\documentstyle[emulapj,epsfig]{article}

\begin{document}

\bigskip

%\magnification=\magstep1
%\baselineskip=12pt \vbadness=10000
%\font\one=cmbx10 scaled\magstep2
\font\two=cmbx10 scaled \magstep1

\title{\bf Unstable Disk Accretion to Magnetized Stars:
First Global 3D MHD Simulations}

\author{Marina M.~Romanova}
\affil{Department of Astronomy, Cornell University, Ithaca, NY
14853-6801; ~ romanova@astro.cornell.edu}

\author{Akshay K.~ Kulkarni}
\affil{Department of Astronomy, Cornell University, Ithaca, NY
14853-6801; ~ akshay@astro.cornell.edu}

\author{Richard V.E.~Lovelace}
\affil{Department of Astronomy and Applied and Engineering Physics,
Cornell University, Ithaca, NY 14853-6801; ~RVL1@cornell.edu }


\begin{abstract}

We report on the first global three-dimensional (3D) MHD simulations
of disk accretion onto a rotating magnetized star through the
Rayleigh-Taylor instability. The star has a dipole field misaligned
relative to the rotation axis by a small angle $\Theta$. Simulations
show that, depending on the accretion rate, a star may be in the stable or
unstable regime of accretion. In the unstable regime, matter
penetrates deep into the magnetosphere through several elongated
``tongues" which deposit matter at random places on the surface of the star,
leading to stochastic light-curves. In the stable
regime, matter accretes in ordered funnel streams and the
light-curves are almost periodic. A star may switch between these
two regimes depending on the accretion rate and may thus show
alternate episodes of ordered pulsations and stochastic light-curves. In the
intermediate regime, both stochastic and ordered pulsations are
observed. For $\Theta > 30^\circ$, the instability is suppressed and stable accretion through
funnel streams dominates.

\end{abstract}

\section{Introduction}

%%%%%%%%%%%%%%%%%%%%%%%%%%%%%%%%%%%%%%%%%%%
\begin{figure*}[t]
\epsscale{1.6} \plotone{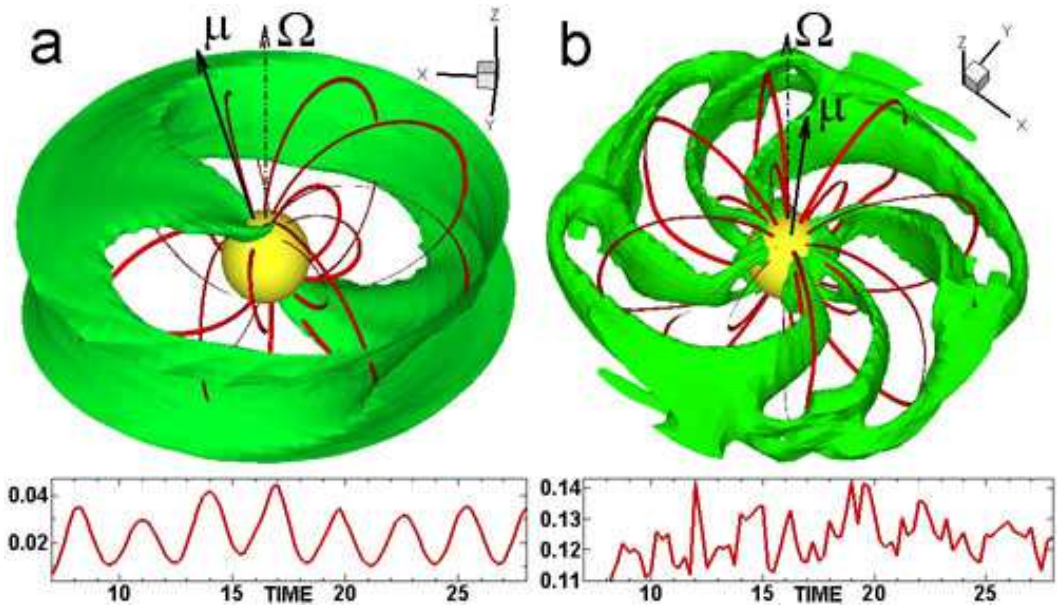}
 \caption{(a) An example of stable accretion. The surface is a constant density
 surface, and the lines are sample magnetic field lines. The magnetic axis is misaligned relative
 to the rotational axis by $\Theta=15^\circ$. (b) An example of unstable accretion for
  $\Theta=5^\circ$.
  The lower panels show light curves from the hot spots.
    Time is measured in orbital periods at $r=1$. Only the inner part of the
    simulation region is shown.} \label{Figure 1}
\end{figure*}

Magnetospheric accretion occurs in different stars, for example in
classical T Tauri stars (CTTSs), which are the progenitors of
Solar-type stars (e.g., Hartmann 1998; Bouvier et al. 2007), in
magnetized white dwarfs in some cataclysmic variables (e.g., Warner
1995; Warner \& Woudt 2002), and in accreting millisecond pulsars,
which are weakly magnetized neutron stars (e.g., van der Klis 2000).
The matter close to the star may either flow around the
magnetosphere, forming ordered funnel streams going towards the
magnetic poles of the star (e.g., Ghosh \& Lamb 1978; Camenzind
1990; K{\"o}nigl 1991), or may accrete directly through the
magnetosphere due to the Rayleigh-Taylor (RT) instability (e.g.,
Arons \& Lea 1976; Elsner \& Lamb 1977; Scharlemann 1978). Earlier
2D and 3D numerical simulations have shown accretion through funnel
streams (Koldoba et al. 2002; Romanova et al. 2002, 2003, 2004).
Recent 3D MHD simulations performed for a {\it wider} range of
parameters have shown that in many cases  the disk-magnetosphere
boundary is RT unstable.

Previous theoretical models and non-global simulations gave useful
but restricted analyses of this problem. Arons and Lea (1976)
investigated magnetospheric accretion  through the RT instability
in the non-rotating case with a spherical accretion geometry. Spruit and
Taam (1990) investigated the stability of an infinitely thin
rigidly rotating disk. The RT instability in magnetized disks
was studied by Kaisig et al. (1992); Spruit et al. (1995); Lubow \&
Spruit (1995). Li and Narayan (2004) investigated disk-magnetosphere
interaction in the case of an infinitely thick disk with a vertical
magnetic field. 2D simulations have been performed by Wang and
Nepveu (1983) and Wang \& Robertson (1985), while Rast{\"a}tter and
Schindler (1999) performed both 2D and 3D simulations, but in a
patch (see also Stone and Gardiner 2007). Such simulations, while
shedding light on many important features of the instability, do not
take into account many factors which are present in global
simulations, such as the possibility of matter flowing through
funnel streams to the magnetic poles of the star.

In this paper we show results from global 3D MHD simulations, where
the simulation region includes the disk and the whole magnetosphere
of the star. This paper summarizes these new results, while in the
following paper (Kulkarni \& Romanova 2008, hereafter - KR08) we
present results for a larger range of parameters.

\section{Model and Results of 3D MHD Simulations}

{\bf 2.1.~Model}. We use a ``cubed sphere" Godunov-type numerical
code (Koldoba et al. 2002) and solve the full set of 3D MHD
equations with initial and boundary conditions similar to those used
in Romanova et al. (2004). Simulations were done for
non-relativistic as well as relativistic neutron stars. We
approximate relativistic effects using the Paczynski-Wiita potential
(Kulkarni \& Romanova 2005). Our model includes viscosity to
regulate the matter flux $\dot M$ through the disk. The viscous
stress in the disk is proportional to the $\alpha$-parameter, which
we varied in the range $\alpha =0.02 - 0.3$. Most of the results
shown here are for the grid $N_r\times N_x\times N_y = 148\times
61\times 61$ in each of the 6 blocks of the ``cubed sphere". Test
simulations for twice as fine a grid and twice as coarse a grid show
that the number of modes does not depend on the grid resolution.

The simulations are done in dimensionless form and are applicable to
stars over a wide range of scales, if the magnetospheric radius
$r_m$ is not very large compared to the radius of the star $R_*$,
$r_m = (4-5) R_*$; $r_m$ is determined by the balance between the
magnetospheric and matter pressure, so that the modified plasma
parameter at the disk-magnetosphere boundary $\beta=(p+\rho
v^2)/(B^2/{8\pi})\approx 1$. The reference units are as follows: the
length $R_0=R_*/0.35$, velocity $v_0=(GM_*/R_0)^{\frac{1}{2}}$,
period $P_0=2\pi R_0/v_0$, density $\rho_0=B_0^2/v_0^2$, magnetic
moment $\mu_0=B_0 R_0^3$, accretion rate $\dot M_0=\rho_0 v_0 R_0^2$
(see KR08 for a complete description of units).
 We show results for a star with a dimensionless magnetic moment $\mu=2$. We
varied the period $P_*$ of the star so that the corotation radius
$r_{cor}=(GM/\Omega_*^2)^{1/3}$ in dimensionless units varied in the
range $r_{cor}=1.2-3$ (with $r_{cor}\approx 2$ in the main case).
The initial disc structure is determined by the initial
disk-corona-magnetic field equilibrium (e.g., Romanova et al. 2002)
and it has the same fiducial density in all runs, so that the
accretion rate in the disk is regulated mainly by the
$\alpha$-parameter of viscosity. The disk is sufficiently large
($R_d\approx 48 R_*$) to supply matter during the whole simulation
run, which lasted up to 50 periods of rotation $P_0$. Most of the
runs  performed were for the small misalignment angle
$\Theta=5^\circ$ which helps excite the perturbations. Runs for
larger $\Theta$ have shown that the instability is present up to
$\Theta \approx 30^\circ$ (see KR08).

\begin{figure*}[t]
\epsscale{1.2} \plotone{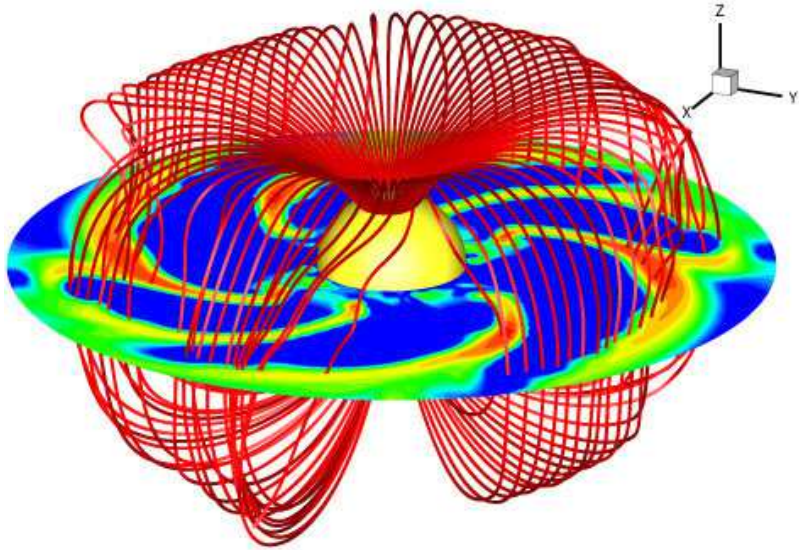} \caption{Penetration of the tongues
through the magnetosphere. The tongues are shown by density contours
in the equatorial plane (the highest density is shown in red, and
the lowest in dark-blue, with a density contrast of about 300).}
\label{Figure 2}
\end{figure*}

\noindent{\bf 2.2.~Stable and unstable regimes of accretion.} In our
earlier work we performed 3D MHD simulations of the
disk-magnetosphere interaction for a variety of misalignment angles
from $\Theta= 0^\circ$ to $\Theta= 90^\circ$, and showed that matter
flows from the inner regions of the disk to the star in symmetric
funnel streams, which are quasi-stable features over long times
(Romanova et al. 2003, 2004). Figure 1a shows an example of
accretion through funnel streams for $\Theta= 15^\circ$. In this
type of flow, matter is lifted above the equatorial plane and then
flows along field lines. The str
eams hit the star near the magnetic
poles, forming ordered hot spots.

\begin{figure*}[t]
\epsscale{1.4}\plotone{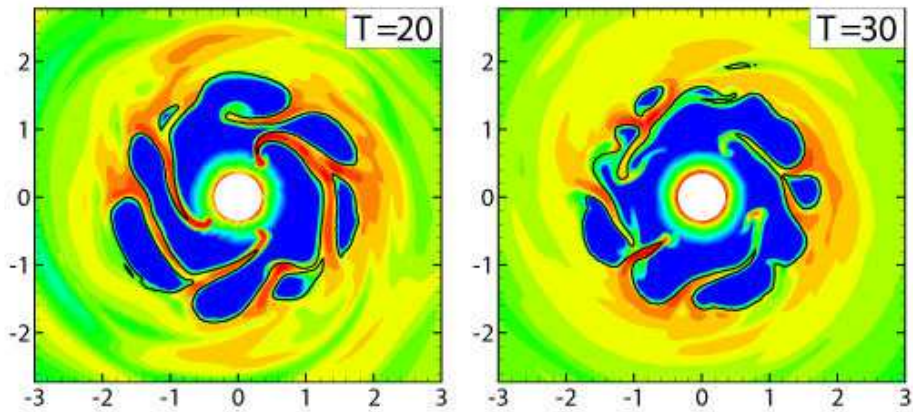}\caption{Density distribution in the
equatorial plane after $20$ and $30$ periods of rotation at $r=1$.
The background shows the density (the colors have the same meaning
as in Figure 2). The black line shows the position of $\beta=1$. The
plots are shown in the reference frame rotating with the star.}
\label{Figure 3}
\end{figure*}

Next we chose a case with a small misalignment angle
($\Theta=5^\circ$) and increased the accretion rate in the disk,
increasing the $\alpha$-parameter. We observed that at a critical
value of $\alpha=0.04$, the instability appears and the accretion
acquires an essentially different pattern: the RT instability
develops at the inner edge of the disk, and matter accretes through
equatorial ``tongues" which penetrate deep into the magnetosphere.
The structure of the tongues is opposite to that of the funnel
streams: they are narrow and tall (see Figure 1b). Such a tongue
shape is determined by the geometry of the magnetic field and has
been predicted by Arons and Lea (1976). Simulations performed for
different parameters and grid resolutions have shown that the number
of growing modes is always small, $m=2-7$ (KR08). The matter
energy-density dominates inside these tongues. Test runs at twice as
large density in the disk have shown that the instability appears
even at $\alpha=0.02$ which proves that the matter accretion rate,
and not $\alpha$, is important. During the interchange process, the
heavy fluid elements of the disk (matter dominated) change positions
with the light fluid elements of the magnetosphere
(magnetically-dominated). The magnetospheric plasma carries
frozen-in magnetic field lines, so that displacing the light fluid
leads to the effect of pushing magnetic field lines aside (see
Figure 2). Equatorial slices of the density distribution for two
moments of time are shown in Figure 3 (see also
http://astro.cornell.edu/us-rus/stereo.htm for animations).

{\bf 2.3.~Light-curves} from the hot spots on the surface of the
star were calculated, to study the difference in observational
properties between the stable and unstable regimes. We assume that
the kinetic energy of the stream is converted at the surface of the
star into isotropic black-body radiation. Integration of radiation
in the direction of the observer is performed (see details in R04).
In the stable regime, the position of the hot spots is almost
constant, so that the light-curve is almost sinusoidal (see Figure
1a, bottom panel). In the strongly unstable regime, sporadically
forming tongues hit the star in random places, so that the light
curve is irregular (see Figure 1b, bottom panel). In the
intermediate cases both funnels and tongues are present, and mixed
light-curves are expected. If a certain number of tongues dominates
(for example, the $m=2$ mode dominates in some cases) then
quasi-periodic oscillations are expected. A more detailed analysis
of the light-curves in different cases will be presented in a future
paper.

\begin{figure*}[t]
\epsscale{1.4}\plotone{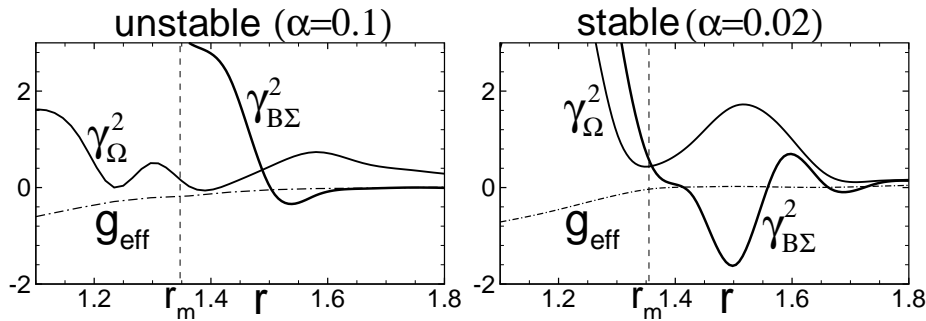}\caption{Parameters corresponding to
the stability criterion by Spruit et al. (1995)
 described in \S 2.4.}
\label{Figure 4}
\end{figure*}

\begin{figure*}[t]
\epsscale{1.4} \plotone{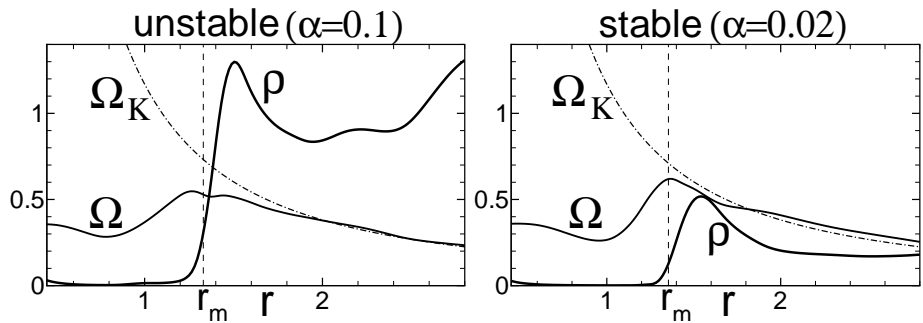}
 \caption{Radial distribution of the density $\rho$, angular velocity $\Omega$
 and Keplerian angular velocity $\Omega_K$ in the equatorial plane.
 The dashed line $r_m$ marks the position
 of the magnetospheric radius where $\beta=1$.}
  \label{Figure 5}
\end{figure*}

{\bf 2.4.~Onset of instability}. We compared our simulations with a
few relevant theoretical approaches. A heavy fluid supported against
gravity by a lighter one is RT unstable unless there is some
opposing force. A homogeneous vertical field at the
disk-magnetosphere boundary does not oppose the growth of
$\phi$-modes (see e.g. Chandrasekhar 1961). A more general
criterion for differentially rotating magnetized accretion disks
indicates that the disk is unstable to excitation of $\phi$-modes if
$\gamma_{B\Sigma}^2 \equiv -g_{eff} d{\rm ln}(\Sigma/B_z)/dr > 2(r_m
d\Omega/dr)^2 \equiv \gamma_\Omega^2$ (Spruit et al. 1995; Spruit \&
Taam 1990; Kaisig et al. 1992; Lubow \& Spruit 1995). Here,
$-g_{eff}=r(\Omega_K^2-\Omega^2)$ is the effective gravity, and
$\Sigma=2 \rho h$ is the surface density. That is, for the
instability to start, the surface density per unit magnetic field
strength $\Sigma/B_z$ should drop off fast enough in the direction
of the effective gravity $(- g_{eff})$, so that the term
$\gamma_{B\Sigma}^2$ is larger than the term associated with the
shear, $\gamma_\Omega^2$, which tends to oppose the instability by
smearing out the perturbations. We calculated the $\phi$-averaged
values of $\gamma_{B\Sigma}^2$ and $\gamma_\Omega^2$ for the stable
and unstable cases (see Figure 4) and compared the positions of the
curves in the inner regions of the disk where the instability
starts, that is at $r \gtrsim r_m$. One can see that in the stable
case ($\alpha=0.02$), $\gamma_{B\Sigma}^2 \lesssim \gamma_\Omega^2$
while in the unstable case ($\alpha=0.1$) $\gamma_{B\Sigma}^2 >>
\gamma_\Omega^2$. Thus the theory makes a reasonably good prediction
for the onset of instability. The term $\gamma_{B\Sigma}^2$ varies
strongly mainly because in the unstable case, a significant amount of
matter accumulates near $r_m$ and the density drop-off is sharper
than in the stable case (see Figure 5).

The RT modes observed in simulations are excited by different
mechanisms. One of them is the small non-axisymmetry associated with
the misalignment of the dipole. Another is the inhomogeneities in
the inner regions of the disk driven by the azimuthal component of
the field trapped inside the inner regions of the disk. The
Kelvin-Helmhotz instability seems to be less significant (see also
Rast{\"a}tter \& Schindler 1999).

{\bf 2.5.~Why low-m modes dominate}. According to the theory of the
RT instability, high-$m$ modes should grow faster than low-$m$ ones
(e.g. Chandrasekhar 1961). In a rotating disk, the shear
$d\Omega/dr$ may efficiently damp high-$m$ modes (e.g., Lubow \&
Spruit 1995) and may be the main reason for damping of the high-$m$
modes in our simulations. Initially, in high-resolution simulations
we see the formation of $m=20-30$ modes, but in one rotation period
they are damped, and only the low-$m$ modes grow. KR08 analyzed the
number of modes based on the Li \& Narayan (2004) criterion and
reached a similar conclusion.

Another possible reason for the suppression of the high-$m$ modes is
the presence of the azimuthal component $B_\phi$ of the field at the
inner edge of the disk which typically constitutes  $\sim (5-30)\%$
of the poloidal component. The damping effect of $B_\phi$ is
expected from theory (Chandrasekhar 1961), and
has been noticed by Wang \& Nepveu (1983) and Rast{\"a}tter
\& Schindler (1999).

\begin{figure*}[t]
\epsscale{1.6} \plotone{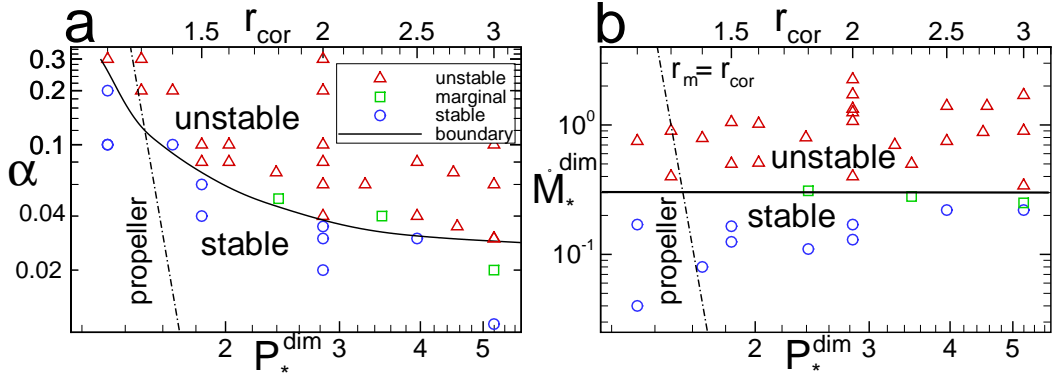}
 \caption{Position of stable (circles), unstable (triangles)
 and marginal (squares) cases in the
 (a) $\alpha$ - $P_*^{\rm dim}$ and (b) $\dot M_*^{\rm dim}$ - $P_*^{\rm dim}$ planes. The
 top horizontal axis shows the corotation radius.
 The solid lines show an
 approximate boundary between stable (below) and unstable (above) runs.
 The dash-dotted line marks the approximate boundary of
 the propeller regime.}
\label{Figure 6}
\end{figure*}

{\bf 2.6.~Boundary between the stable and unstable regimes}. We
performed multiple simulation runs for identical initial density
distributions in the disk but different viscosity parameters
$\alpha$ and dimensionless periods of the star $P_*^{dim}$ (these
bulk simulations were done for a grid $72x31^2$). The symbols in
Figure 6a correspond to different runs. One can see that there is a
boundary between the stable and unstable cases with unstable cases
corresponding to higher $\alpha$. For similar initial disks and
accretion rates through the disks, $\dot M$ is proportional to
$\alpha$, and thus the instability occurs at high enough accretion
rates through the disk. The boundary is higher at smaller periods
$P_*^{dim}$, because at faster rotation of the star the effective
gravity becomes less negative and thus less matter may accumulate at
the inner edge of the disk. That is why a larger $\alpha$ is
required to bring more matter to the inner edge of the disk to
satisfy the instability criterion. Cases with periods
$P_*^{dim}\approx 2.7 - 3$ correspond approximately to the
rotational equilibrium state (Long et al. 2005). Stars with smaller
periods spin down and, at $P_*^{dim}\lesssim 1.8$, enter the
``propeller" regime, where $r_{cor} < r_m$ (e.g. Illarionov \&
Sunyaev 1975; Lovelace et al. 1999; Rappaport et al. 2004). In this
regime a star tends to expel a significant part of incoming matter
(Romanova et al. 2005; Ustyugova et al. 2006). At high enough
$\alpha$ we observed events of instability even in this regime (see
also Wang \& Robertson 1985), but the simulation runs were brief,
and this region requires further analysis.

We calculated the accretion rate onto the {\it surface of the star}
$\dot M_*^{dim}$ for the above runs. The boundary between stable and
unstable accretion $\dot M_*^{dim}\approx 0.3$  seems to be almost
independent of $P_*$, which is probably a result of the combination of the lower
disk density due to the propeller effect, and higher $\alpha$. This
boundary is approximate, with intermediate cases above and below it
in which only marginal instability is observed. In this transition
region, both funnels and tongues are present. Higher up, for $\dot
M_*^{dim}> 0.5$, the instability is strong and accretion through
tongues dominates. We note that at low enough accretion rates
through the disk even stable accretion may shut-off, because there
is insufficient pressure gradient at the disk-magnetosphere boundary
to form magnetospheric funnel streams.

{\bf 2.7.~Application to different magnetized stars.} The results of
our simulations can be applied to a variety of magnetized stars with
small magnetospheres, $r_m\approx (4-5) R_*$, including CTTSs,
weakly magnetized white dwarfs (WDs) in cataclysmic variables and
neutron stars (NSs) in LMXBs. The dividing line in Figure 6b
corresponds to $\dot M_*^{dim}\approx 0.3$. Below we show the
accretion rate corresponding to this boundary in dimensional units
for different types of stars:

$$\noindent {\rm CTTSs:}~~ {\dot M_*}^{\rm cr}\approx 2.1\times 10^{-8}
({{B_*}/{10^3 {\rm G}}})^{2} r ^{\frac{5}{2}}
m^{-\frac{1}{2}}{M_\odot}/{\rm yr},$$

$$\noindent{\rm WDs:}~~~{\dot M_*}^{\rm cr}\approx
1.4\times 10^{-8}({B_*}/{10^6 {\rm G}})^{2} r^{\frac{5}{2}}
m^{-\frac{1}{2}} {M_\odot}/{\rm yr},$$

$$\noindent{\rm NSs:}~~~{\dot M_*}^{\rm cr}\approx 2.2\times 10^{-9}
({B_*}/{10^9 {\rm G}})^{2} r^{\frac{5}{2}} m^{-\frac{1}{2}}
{M_\odot}/{\rm yr}.$$

\noindent The period in dimensional units is:

$$ P_*=A_p P_*^{\rm dim} m ^{-\frac{1}{2}}r^{\frac{3}{2}},$$

\noindent where scaling factors for different stars are:

\noindent {\bf CTTSs:} $A_p\approx 1.78~ {\rm days}$,
$r=R_*/2R_\odot$, $m=M_*/0.8 M_\odot$,

\noindent{\bf WDs:} $A_p\approx 29.4 ~{\rm sec}$, $r=R_*/5\times
10^8 ~{\rm cm}$, $m=M_*/1 M_\odot$,

\noindent{\bf NSs:} $A_p\approx 2.22~{\rm ms}$, $r=R_*/10^6~{\rm
cm}$, $m=M_*/1.4 M_\odot$.

 Note that the dimensional ${\dot M}_*^{cr}$ increases with
the magnetic field as $B_*^2$ so that the boundary will be higher
for higher $B_*$. This reflects the fact that we have fixed the
dimensionless magnetic moment $\mu$, and thus consider stars with an
approximately fixed ratio $r_m/R_*\approx 4-5$. We performed another
set of runs for smaller magnetospheres (smaller $r_m/R_*$ for
$\mu=0.5$) and observed that the dimensionless boundary corresponds
to a smaller $\dot M_*^{dim}$ (see KR08). Note also that all the
above runs were done for $\Theta=5^\circ$. We expect that for larger
$\Theta$ the boundary will move up, because funnel accretion is more
favorable at larger $\Theta$.

\section{Observational Consequences}

The existence of two regimes of accretion changes our understanding
of accreting magnetized stars and their observational properties:
the lack of a periodic signal in the light curve does not rule out
accretion and an ordered magnetic field. For example, the random
light-curves in many CTTSs or dwarf novae do not preclude an ordered
stellar field, if the star is in the unstable regime. Depending on
the accretion rate, episodes of stable and unstable accretion may
alternate, so that periods with pulsations may be {\it intermittent}
and may be followed by periods with no pulsations. Recently a number
of intermittent accreting millisecond pulsars have been discovered
(e.g., Kaaret et al. 2006; Altamirano et al. 2007; Galloway et al.
2007; Gavriil et al. 2007). Although the reason for this behavior is
not yet understood, we note that in, for example, HETE
J1900.1--2455, the pulsations disappeared when the overall X-ray
flux increased (Kaaret et al. 2006) which may be an example of a
transition to the unstable accretion regime.

In some cases a definite number of tongues may dominate, which may
lead to quasi-periodic oscillations in the light curves (Li \&
Narayan 2004). These oscillations may be candidates for one of the QPOs
observed in Type II (accretion-driven) bursts in LMXBs.
Comptonization of photons by high-energy electrons may lead to only
a small departure of the light-curve from the thermal ones obtained
using the approximation of isotropic black-body radiation (Poutanen
\& Gierli\'nski 2003), so that the QPO features may survive
Comptonization (see, however, Titarchuk et al. 2007).

In the case of young stars surrounded by gas/dust disks where
planets are forming and migrating inward, the unstable tongues may
support inward migration, so that the magnetospheric gap, which can
halt migration (e.g. Lin et al. 1996; Romanova \& Lovelace 2006;
Papaloizou 2007), may form only in the state of stable accretion.

\acknowledgments We thank the anonymous referee for valuable
suggestions which improved the paper. We thank Drs. A. Koldoba and
G. Ustyugova for their contribution to the code development and Drs.
D. Altamirano, L. Hillenbrand, P. Kaaret, and C. Thompson for
discussions. NASA provided high-performance computational facilities
for this work. The research was partially supported by the NSF
grants AST-0507760 and AST-0607135, and the NASA grants NNG05GG77G
and NNG05GL49G.

\end{document}